\documentclass[12pt]{article}

\usepackage{graphics}
\usepackage{amsmath}
\usepackage{amsfonts}
\usepackage{amssymb}

\title{Carlo Cercignani's  Interests for  the Foundations of Physics}

\author{Luigi Galgani\thanks{Universit\`a di Milano, Dipartimento di Matematica,
         Via Saldini 50, 20133 Milano,
         Italy }}

\begin{document}

\maketitle

\begin{abstract}
Carlo Cercignani was known all over the world for his works
  on the Boltzmann equation and on kinetic theory. There was however
  another aspect of his scientific life, which is not much known.
Namely,  his interest for the foundations of physics, in particular
for  the possibility  of understanding quantum mechanics through
classical mechanics, which he shared with several people in Milan. A
review of such researches is given here, together with some personal
recollections of him.
\end{abstract}

\section{Introduction}
Carlo Cercignani was well known all over the world as one of  the
greatest experts in the problems related to the Boltzmann equation and
more in general
to kinetic theory, to
which most  of his scientific life was devoted. Very little known is
instead  his interest
for the foundations of physics, especially the relations between
classical and quantum mechanics. Two papers have a special interest in
this connection; the
first one \cite{zpe} of the year 1972, by the title 
\emph{``Zero--point energy in classical
  non--linear mechanics''},    in
collaboration with Antonio Scotti and me, and  the second one
\cite{cerci} of the year 1998,
 by the title
\emph{``On a nonquantum derivation of Planck's distribution law''}. 
 His interest is also witnessed by another  paper on the same  
subject, namely, \emph{``Quantization \`a
la Nernst for the plane rotator''} \cite{montaldi}. A further paper 
\cite{universal}, with G. Benettin,
A. Giorgilli and me (see also \cite{feig2}), will also be mentioned . 
It reports on the 
discovery of the analog of the universal Feigenbaum number for
Hamiltonian systems, and  its hidden connection with the general problems
discussed here will be indicated.
 
To say the core of the things  in a few words, Carlo Cercignani shared with our
 group forty years ago, and continued to share along all of his
 life, the idea that the last word has not yet been said about the 
relations between classical and quantum mechanics. Perhaps new
insights might  come from the theory of classical dynamical
 systems, after the great impulse it received since  the year
 1954, both from   the mathematical  work of
 Kolmogorov (followed by those of Arnold, Moser and Nekhoroshev), and the
 physical work of  Fermi, Pasta and  Ulam.

A quick illustration of such works of Carlo Cercignani will be given
here, after a recollection of the human and  scientific athmosphere 
in which they were conceived. A short outline will also be given of
the present state of the art in the field. Finally, some
recollections will be given concerning certain personal aspects of his
life.

\section{The Fermi Pasta Ulam problem and the foundations of physics
 in Milan in the years sixties}
Things went as follows. Carlo had graduated in physics at Milan
 in the year
1961, and two years later also in mathematics,  and had a great
reputation among young people working
around the Institute of Physics, at via Celoria 16. Actually such a
reputation  even went back to
the times when he was a student  and he was considered  a kind of a genius.
 At the end of the years sixties he was working on some
applications of hydrodynamics and 
kinetic theory, but from time to time he used to come to the
Institute, where for instance he 
gave informal lectures on various sophisticated mathematical problems.
There was a rather large  group of young people working at the
Institute or somehow associated with it, 
and the general athmosphere was very beautiful: one used to
meet in the evenings, to go to the Alps in the week--ends, .... For
what concerns  research, in Milan there was a certain tradition in
studies concerning foundational problems, going back to Caldirola
and his pupils Loinger, Bocchieri, Prosperi, Scotti,  Ghirardi, Lanz
and others.
Caldirola liked to mention that on several occasions he had the opportunity to
take part in discussions within the Fermi group in Rome. Moreover
Pauli had come several times to Milan, my tutor Montaldi had been
several years in Munich with Heisenberg, Bruno Bertotti  from
Pavia had been with Schroedinger in Dublin, and Rosenfeld was
considered as some kind of sponsor of the Milan group. For example, when
Scotti and me  first  became aware of  KAM theory (that in Italy
absolutely no
one knew in those years), I was requested to write down a short review 
that was sent to Rosenfeld. An enlarged version was later published,
together with Antonio Scotti \cite{gs}, for La Rivista del Nuovo Cimento.

One day Loinger  discovered\footnote{I do not know in which
  way. Perhaps, through the work of Izrailev and
Chirikov (1966), although  I came to know of such a paper
 only  much later.}  the work of Fermi, Pasta and Ulam (FPU), in which 
the relevance of dynamics  in connection with the equipartition
principle was pointed out \cite{fpu}. This, as everyone knows, is the key point  
where quantum
mechanics had its origin by Planck, on October 19, 1900. One is
concerned with the distribution of energy among a system of
oscillators of different frequencies, and classical mechanics, through
the Maxwell--Boltzmann distribution or the Gibbs ensemble which should
govern equilibrium statistical mechanics, predicts that at equilibrium
at a temperature $T$ all oscillators should have the same mean energy 
$U(\nu,T)=kT$ ($k$ being the Boltzmann constant), irrespective of
their frequencies $\nu$. This is what equipartition actually
means. Phenomenologically, instead, one finds that the high
frequencies have very little energy, actually one that decreases
exponentially fast with  frequency. The distribution of energy is
described by Planck's law 
$$
U(\nu,T)=\frac {h\nu}{\exp(h\nu/kT)-1}\ ,
$$
which was just invented by
a skillful fit to the data with the introduction of Planck's constant
$h$, and quantum mechanics with its energy
levels $E_n=nh\nu$ was  just built above it.

Thus, the conflict between classical mechanics with its equipartition,
and quantum mechanics which just by definition is Planck's law, is
patent, and about say 99 percent of scientists decrete that there is
no problem: classical mechanics is wrong, being valid just in some
limit situations, typically for low frequencies $\nu$ or high temperatures
$T$, actually, in the limit $h\nu/kT\ll 1$, where  Planck's law meets 
 equipartition.

However, there are a few scientists who feel that  things are subtler,
and there still is a delicate dynamical aspect of the problem that
should be settled. This actually is the point that Fermi (with Pasta and 
Ulam) made in  what probably was  the last work of his life. 
Notice that
he had addressed the problem already in the year 1923, by
elaborating on a theorem of Poincar\'e about the number of  integrals of motion
of a generic Hamiltonian system, which is usually interpreted as
supporting the common opinion that classical mechanics implies
equipartition also in a dynamical sense. But Fermi was not content
with this,
and when for the first time he had available a computer that might
allow him to check  conjectures by ``numerical experiments'',
he made a dynamical check of equipartition. He considered the system
of oscillators which correspond (as normal modes, in the familiar way)
to a system of point particles on a line with nearest neighbour
slightly  nonlinear forces. For initial data with energy given just to
one low--frequency oscillator, or to a few low--frequency ones, 
energy was expected to flow up,
equally distributing itself among all oscillators, as required by the
equipartition principle as equilibrium is attained. Instead, the paradoxical
result was found that, up to the available computational time,
a state of apparent  equilibrium was attained, in which
energy was shared only among a small packet of low--frequency
oscillators, while the high--frequency ones essentially remained
deprived of any energy. Such a ``final'' state appeared to
be a stationay one, which did not move at all. According to Fermi,
this was some new, relevant  discovery. For the figures, see the
original paper, or for example the review \cite{noigiovanni}. 

So,  the FPU ``paradox'' saw the light, and it made patent 
the deep role played by  dynamics in statistical mechanics. By the
way, just in the same year 1954 Kolmogorov had proven his celebrated
theorem on invariant tori, on which Moser and Arnold came back in the
years 1961, 1962. Thus  some kind of a new era in
the mathematical aspects of perturbation theory had started, 
and at the same time a new 
popular acquaintance with
the coexistence of ordered and chaotic motions took place, especially
through a diffuse   availability of computers.. 

Now, which had been the reaction of the scientific community to the
FPU paradox? In the year 1965 \cite{zab}
Kruskal, Zaburksy   and their collaborators had taken up a
certain mathematical side of the problem, paving the road to solitons and
integrable partial differential equations, but had said nothing on the
physical side of the FPU problem. However, such works impressed Carlo
very much, and he studied them carefully. What came out is a review by
the title \emph{``Solitons. Theory and application''} \cite{solitoni}
that he wrote for La Rivista del Nuovo Cimento.

The physical relevance  was instead immediately proclaimed   one
year later by
Izrailev and Chirikov \cite{izc}, who  completely understood
that   a fundamental problem had been opened. They also
were so clever as to indicate a possible way out. 
Let us ponder on this. First of all they made a crucial discovery. 
Namely,  if one
repeats the FPU ``experiment'' for initial data exactly as
FPU, but at a  high enough energy, then 
the FPU paradox disappears,
because    equipartition is attained very quickly. In other words, there
exists a kind of critical energy $E_c$, and the paradox only occurs
for an energy $E<E_c$. Then they also imagined a way in which
the paradox could be completely eliminated for macroscopic systems, i.e.,
 for systems with a very large number $N$ of constituents
(formally, in the so--called thermodynamic limit  
 $N\to \infty$,  for fixed values of the
specific energy $E/N$ and of the specific ``volume'' $L/N$, if $L$ is
the size of the FPU system). The very
simple idea is that the critical specific energy $\epsilon_c=E_c/N$ vanishes in
the thermodynamic limit. Then, for any macroscopic system, which by
definition has specific energy $\epsilon=E/N >0$, one would have  
$\epsilon>\epsilon_c$, and the
``ordered'' FPU--like states violating equipartition would not exist at
all. With such a conjecture the problem of eliminating the FPU paradox
then became a mathematical one, namely, to prove or indicate in any
possible way, that the  energy threshold $E_c$ does not grow at all or
grows less than  $N$ for increasing $N$.

This was the situation when Loinger discovered the FPU problem. Now,
at the end of the years sixties Scotti  had come back to Milan from
California where, working in  high energy phyisics, he
had  become acquainted with the use of computers. So Loinger
involved him and Bocchieri, and they repeated the computations for a
FPU--type  model  with realistic intermolecular
forces, given by  the so--called Lennard--Jones potential 
$$
V(r)=4V_0
\big[(\sigma/r)^{12}-(\sigma/r)^6\big]\ ,
$$ 
which involves two parameters $V_0$ and
  $\sigma$. The only other parameter  entering the model was the mass $m$ 
of the particles. They too confirmed the
existence of an energy threshold $E_c$, but their new contribution
was an indication, based on their numerical results, 
that the specific threshold $\epsilon_c=E_c/N$ may
not vanish with increasing $N$ \cite{boc}. 
This result was perhaps even expected 
because, when shortly later  Loinger gave a talk on the general
problem of the relations between classical
and quantum mechanics, he explicitly said he liked that result. Most
people in the audience  judged him as crazy. I, instead, was
fascinated.

At that moment I entered the game. Since I was already working with
Scotti on equilibrium statistical mechanics, and really had begun to
share my life with him,  I started helping him in
some further computations he was performing on the FPU problem. We
were impressed by the fact that the ``final'' distribution of energy among
the oscillators was decreasing exponentially fast with frequency, and
just had the crazy idea to fit it to a Planck--like law. The fit turned out
to be rather good, but what really struck us as something incredible
and almost absurd, was what occurred with the  free parameter
introduced in the fit. Indeed we had chosen a
Planck--like distribution in which the only  free parameter was an
action $A$ taking the place of Planck's constant $h$, and the fit gave for
$A$ a value rather near to $h$. It took us perhaps one
month  to really understand what had occurred. If in a
classical mechanical model one determines an action $A$, then this has
to be
a pure number, say $\alpha$, times  some  ``natural'' action which is
defined in terms of the parameters entering   the model. In our case
they were the parameters $V_0$, $\sigma$ and $m$ previously mentioned,
and so standard dimensional analysis gives
$$
A=\alpha \sqrt{m V_0}\ \sigma\ .
$$
Now, from the relevant textbooks on molecular physics one finds that
for the rare gases one has
$$
\sqrt{m V_0}\ \sigma=2Z\hbar
$$
where $Z$ is the atomic number and $\hbar=h/2\pi$. So $h$ had been
introduced into the classical FPU model from outside,
 through the molecular  parameters of the Lennard--Jones potential. 
Moreover we had actually
worked, as previously Bocchieri Scotti and Loinger, with the
parameters of Argon,
and by chance the value of the pure number $\alpha$ was such that,
combined with the values of the molecular parameters of Argon, 
it produces $A\simeq \hbar$.

In any case the FPU paradox was  thus enhanced, and Planck's law
appeared to enter classical physics, at those low temperatures for
which equipartition is not  attained dynamically. This remark
was published  in the Physical Review
Letters \cite{prl}. Actually  Bocchieri and Loinger did not dare to sign the
paper, and Scotti and me were   left as the only two authors.
The reactions in the scientific community were in general  not very good.

\section{Enters Carlo. The 1972 CGS work}
Here, Carlo enters the game. One day I was explaining these things on
a blackboard to Maria Marinaro, a physicist from Naples whom I had 
met in London a little before, at a conference where I had presented our
results.  While I was discussing
with her, Carlo showed up at the door of the room.  All
doors were left opened in those years, and everyone passing through could freely join
a discussion. I still remember Carlo on that occasion, elegant as
usual in his blue
suit,  leaning casually  against the door jamb, and listening with great
attention. After perhaps ten minutes he left. But a few minutes later
he came back, and
told me that he too had happened to think of the relations between
classical and quantum mechanics, although he never had 
heard  of the FPU
problem. So we arranged that we should talk about it, together with
Scotti, and so we did in a few days.

What he had in mind is the role of zero--point energy.  This, as
Planck himself says in the preface to his book on the quantum theory
of radiation, is the quintessence itself  of quantum mechanics. 
As is  well known, with his ``second theory'' of the year 1911 Planck
had suggested   that each 
oscillator should possess, in addition to the thermal energy $U(\nu, T)$
 given by the standard Planck's formula, also an energy $h\nu/2$. So the total
energy should be $E(\nu,T)=U(\nu,T)+h\nu/2$. i.e., 
$$
E(\nu,T)=\frac {h\nu}{\exp(h\nu/kT)-1}+\frac 12 h\nu .
$$
At zero temperature the ``thermal part'' $U$ vanishes, and only the
``zero--point'' energy (\emph{nullspunkt} means zero temperature in german)
remains. This  energy  of a non thermal nature is known to produce
mechanical effects, such as the so--called Casimir effect.  So one has
on the one hand an energy of disordered or  chaotic type,  and on the
other one an energy of ordered
type, as was particularly stressed by Nernst in the year 1916.
 But at that time no one of us was aware of these
features concerning  such two  types of energy, and the main idea of Carlo
was precisely to introduce such notions in the FPU frame. 
In such a way he was led to conceive that one should go beyond the
conception of a global energy threshold of the system,  introduced
by  Bocchieri Scotti and Loinger.
 He instead insisted on the fact that each frequency should
present a separate energy threshold, which should be identified with
Planck's  zero--point energy.

Actually, the idea that zero--point energy should play the role of
some threshold between ordered and chaotic motions can even be found in the
very original paper of Planck of the year 1911, although Carlo only knew what Planck
says in his book. Indeed, in Planck's paper the conception is advanced
that, when an oscillator absorbs energy starting from a state of no energy,
the process is ``regular'' until the oscillator  reaches the 
energy\footnote{Both Planck and Nernst propose that the value of the
  zero--point energy should be  $h\nu$ rather than $h\nu/2$.} $h\nu$.
At that moment, Planck imagines,   the process becomes dynamically 
so complicated and
unstable that a purely dynamical description becomes impossible, and
some probabilistic considerations become necessary. So the oscillator
has a certain probability of having its energy  increased, and the
complementary probability to lose energy falling back to the state of
no energy.

In order to check the reliability of the  conjecture advanced by Carlo,
that the threshold of the FPU model could correspond to zero--point
energy, what we did  was just checking that the total zero--point energy 
$(1/2)\sum_ih\nu_i$ turns out to be of the same order of magnitude of
the total critical energy $E_c$ discovered by Bocchieri  Loinger and
Scotti, and this actually turned out to be the case. Some days later
  I wrote a paper in a few hours, Carlo and Tonino Scotti added a
few remarks, and the paper was sent and published in Physics Letters
\cite{zpe}.

One might mention that no one of  us even knew of the 
paper of Izrailev and Chirikov, which we came to know much later. It
actually turns out that the idea that
there should exist an energy threshold for each frequency, leading from
ordered to ``chaotic motions'',  actually is the main point of the
russian authors. But Carlo's  intention was exactly  the opposite one.
 Indeed Izrailev and Chirikov wanted to use such a feature as
a means to  eliminate the physical relevance of the  FPU problem,
just by trying to prove that the thresholds vanish in the thermodynamic
limit. Instead, Carlo's idea was that such thresholds should be the
mathematical counterpart of a physical phenomenon which reflects a ``reality'',
namely, the coexistence of ordered and  chaotic motions, or the
coexistence of a thermal and a non thermal energy. And this can only
obtain if the thresholds do not vanish for macroscopic systems. 

In conclusion,  Carlo
invented by himself the idea that 
\begin{enumerate}
\item 
 a  threshold for each frequency
should exist, 
\item
 it should persist in the thermodynamic limit, and
\item
 it should play the role of the door through which quantum
mechanics may be ``explained'' by classical mechanics. 
\end{enumerate}

Carlo, who is
known everywhere for his  great technical abilities, his mental
powers  (for example, he could recite  all of
Dante's Divine Comedy)\footnote{I once was joking with him about
  this, telling him that according to some people he could do it
  both forward and backward. And
  he replied: `` saying backward, do you mean verse by
  verse, or word by word?''}  and
his great skill, was in that case able to produce an incredible
free invention, just based on pure intuition or phantasy, and even
dared to 
 involve
himself in a game that most in the scientific community consider with
great skepticism, or suspicion, to say the least.

\section{Present state of the FPU problem}
After such a proposal concerning the FPU problem and quantum mechanics,
that involved Carlo, Tonino Scotti and me, in a 
sense the community of scientists working on the foundational aspects of
the FPU problem,  turned out to be divided into two groups; those
that would like to prove that the thresholds disappear in the
thermodynaimc limit, and those  that would like to prove the
contrary. There are here two extremely difficult problems. The first one
is to invent a concrete mathematical position of the problem which
suits the physical interpretation. The second one is of an
analytic nature, and consists in the corresponding  technical proof of
an actual theorem, in the strict mathematical sense. 

One might ask  whether the situation has presently  been clarified, after
so many years. The answer is still negative. Some substantial
progress has however
been made, and I will try now to mention a few  steps
that appear particularly relevant to me. More details on certain
aspects of the problem can be found in a review that I wrote with
Giancarlo Benettin, Andrea Carati and Antonio Giorgilli 
\cite{noigiovanni} on the
occasion of a conference on the FPU problem organized by Giovanni
Gallavotti in Rome. Other points of view illustrated at that
conference may be found in  the book \cite{giovanni}. 

The first progress was   a paper written in the year 1982
 by a group of persons around Giorgio Parisi \cite{fucito}, 
where it was proposed
 that the ```final'' FPU state actually is an apparent equilibrium
 state, i.e., a  metaequilibrium one, which on a second extremely long
 time scale might approach the final standard equilibrium. Moreover, a
 very beautiful mechanism was conceived to explain analytically how a
 quick approach is made to the intermediate  metaequilibrium state.  A
 visual numerical exhibition  of  such a passage from metaequilibrium to
 equilibrium was then given by Antonio Giorgilli, Luisa Berchialla and
 me \cite{luisa}. Furthermore the mechanism for the quick approach to
 metaequilibrium proposed by the
 group around Parisi was investigated by Dario Bambusi and Antonio Ponno
 \cite{dario} in the spirit of the old works of 
 Kruskal and his collaborators, and a new
 path was thus opened to the use of PDE's as normal forms for
 FPU--like systems.

The second progress concerns the problem of how should one 
formulate an analog of the FPU problem if one
considers initial data of ``generic type'', rather that of the special
FPU type in which only low--frequency oscillators or just some few
oscillators are initially excited. Indeed in the latter case the
attainment of  equipartition is a mark of having reached
equilibrium. Instead, for generic initial data,  one is in
presence of equipartition already from the
start, and so one has to invent some new  way of
establishing whether some analog of the FPU phenomenon even exists
in that case. The
answer that was proposed \cite{fdt}\cite{simone} is that one should 
look at the problem from
the point of view of the Fluctuation Dissipation Theorem, which
concerns the exchange of energy between two interacting systems, 
for example a FPU system in contact  with a heat 
reservoir. In such a case, equilibrium is attained when the
time autocorrelation of the FPU system's energy  has decayed to zero, and an
analog of the FPU phenomenon occurs if the autocorrelation decays to
some nonvanishing value, and moreover remains stabilized at that
value. Physically, this corresponds to the fact that the measured
specific heat of the FPU system is not the classical canonical one,
but a smaller one, as qualitatively occurs in quantum mechanics.

The third progress consists of an extension of classical perturbation theory   
to the thermodynamic limit ($N$ tending to infinity with 
nonvanishing specific energy  $E/N$ and specific volume $L/N$). Classical
methods do not work in such a case. The great advance was performed by
Andrea Carati \cite{andrea}, who showed how results independent of $N$ 
are obtained
if one weakens the requirements, and just formulates  the problem in a
probabilistic frame by introducing $L_2$ norms with respect to
Gibbs measure, in place of the standard sup norm. 
In such a way he has
obtained quite recently, together with his pupil  Alberto Maiocchi,  a
Nekhoroshev--like theorem in the thermodynamic limit
 for the so--called $\phi^4$  model 
(a simple variant of the FPU one)\cite{alberto}.

The big problem which remains open is whether stability  results 
of the type just mentioned, namely, in the thermodynamic limit, 
apply also to FPU models in dimension  two or three. Indeed, some
numerical computations performed by Giancarlo Benettin 
\cite{gianca} \cite{gianca2} appear to
suggest that, in passing from dimension one to dimension two, the FPU systems 
tend to become more chaotic, so that the thresholds might
disappear. 

So, we are now, in the year 2011, more or less in the
same situation as in the year 1972, when the  proposals of Izrailev
and Chirikov and of Cercignani--Galgani--Scotti stood one against 
the other. The situation did however change because,  with the 
analytic progress  recently accomplished  in perturbation theory in the
thermodynamic limit one may be confident that  the
problem will be finally settled analytically, in favour of the one
thesis or the other. I obviously
hope   (and I'm sure Carlo too  would)
that the thresholds will be proven to persist in the thermodinamic
limit, even for dimensions two and three. If not, it will in any case
be true that  our idea proved to be helpful in stimulating deep analytical
results, and all of us  will content ourselves with applying   the
old italian adagio \emph{``Se non \`e vera, \`e ben
  trovata.''}\footnote{If it is not true,  it is at least a beautiful
  invention.}

\section{A strange work on universal numbers}
 Let us now come to the work \cite{universal}. This deals with the well
 known discovery of Feigenbaum concerning universal numbers in
 dynamical systems. By studying some  mappings depending on a
 parameter (say $\mu$) he found that a sequence of bifurcations occur, 
of ``period doubling'' type: at $\mu_1$, say,  a fixed point passes from
stable to unstable and  a periodic orbit of period two shows up, then
at $\mu_2$ that orbit becomes unstable and an orbit of period  four
shows up, and so on. The sequence $\mu_n$ has a limit, and  he
observed that the ratio $(\mu_{n+1}-\mu_n)/(\mu_{n}-\mu_{n-1})$ too
tends to a limit which is universal, i.e., does not depend on the
mapping. The latter limit is the universal Feigenbaum number
$4.669..\, $. 
In our paper we were the first in the world to point out  that for
conservative mappings (i.e., for the Hamiltonian equations of motion of 
mechanics) the corresponding number does not coincide with the
Feigenbaum one, still being universal for the subclass of Hamiltonian
systems. Such a number is $8.721...\, $. 

But how did we happen to be interested in this problem, and which are
the relations with the general problems discussed here? Things went as
follows.
In the summer  1980 I was attending a conference in Pojana Brasov
(Romania), and an
afternoon I decided to go for a walk. But Herbert Spohn met me, and
insisted that I should listen to a talk by Eckmann, which  he expected
to be very interesting. So I went. The talk  was very interesting indeed, and
was a general review about the Feigenbaum matter, of which I knew
nothing. 
As everything
was concerned with dissipative systems, I asked the speaker which was
the situatiom concerning Hamiltonian systems, and  he told me that a
work had been published containing the answer, namely, that the same
phenomenon had been observed in Hamiltonian systems too, and with the
same Feigengbaum number. 

The reason why I was interested in universal numbers is again the
relation  between quantum and classical mechanics, in connention with
the celebrated pure number 137. This had been introduced by Sommerfeld 
in connection
with the fine structure constant, and  essentially goes back again to
Planck's constant $\hbar$. Indeed $\hbar$ is an action, and if one
looks for a classical analog of it in the frame of electrodynamics one
meets with the action $e^2/c$ where $e$ is the electron charge and $c$
the speed of light. So $\hbar$ is a multiple of $e^2/c$ and the
factor is just $137.035999679...$. In other words there exists the ``fine
structure constant''  $\alpha=e^2/\hbar c$, a pure number, 
and one has $\alpha\simeq 1/137$.

This is a point on which I happened to  have infinitely many conversations with
Carlo and Tonino Scotti, and many times we  indulged in
imagining how that magic number 137  might be extracted from some hidden
dynamical property of Hamiltonian systems. All this came to my mind
during the talk of Eckmann  in Pojana Brasov. So when I came back to Milan I
discussed this new perspective with Carlo, and we decided to make an
attemp. Actually the work was done by my two jewels, Giancarlo
Benettin and Antonio Giorgilli who, independently in Padua and in
Milan, conceived some numerical method to look for fixed points,
especially suited for conservative, as opposed to dissipative,
systems. Actually both of them found out that the published results 
were incorrect, that Hamiltonian systems indeed have a Feigenbaum--type
number, the value of which however  is $8.721...$, different
from the Feigenbaum one. But also different from  $137$ or
$1/137$. We were disappointed, but in any case a new original result
in Hamiltonian systems had been discovered (see also \cite{feig2}). 
This is indeed the moral
of all the story. We are looking for something which is considered
 crazy, and discover different, but in any case new,
things. Somehow  as Colombo, who wanted to go to India, and discovered America.

\section{Carlo comes back to Planck's law: the Foundations of Physics
  1998  paper}
For a long time Carlo was mostly involved in his main research themes
on  kinetic theory, while in my group we were continuing studying
the mathematical aspects of perturbation theory and of the FPU
problem. We also had started a deep study, which still continues in the
present days, on the dynamical aspects of electron
theory when radiation reaction and retardation are taken into account
(see for example \cite{nc}\cite{massimo} and \cite{plasmi}). 

For what concerns a
possible deduction of Planck's law in a classical frame I had done a
consistent progress (see the papers \cite{dednernst} \cite{dednernst2}
and the paper \cite{dednernst3} with Giancarlo Benettin). 
This passed through a deep 
understanding (which took about ten years) of the 
paper of Nernst of the year 1916, by the title \emph{``On an attempt to
  return from quantum mechanics  to the assumption of  continuous 
variations of energy''} \cite{nernst}, the main
point of which was the conception that Planck's law may be consistent
with equipartition. This is  another way -- possibly the first one,
historically -- of dealing with   the idea, 
already mentioned in
this paper, of  the Fluctuation Dissipation Theorem.
Namely, that Gibbs' law, with the corresponding equipartition, just
plays the role of a measure for the initial data, whereas the actually
exchanged energy depends on dynamics, in particular on the existence
of an energy 
threshold. Nernst was indeed speaking in terms of a
threshold between
ordered  (\emph{geordnete}) and disorderer  (\emph{ungeordnete}) 
motions. So I had given  some interpretation  of  Nernst's deduction
of Planck's law,
which makes an explicit reference to the energy thresholds, the role of which
had  been particularly stressed by Carlo. In this connection,  the
paper \cite{montaldi} too was written.

Much  time passed, and one day Carlo showed me an article 
he had written  on
that subject, and after some adjustments he published it in the
Foundarions of Physics Letters \cite{cerci}. I actually never fully 
understood the
new point he was making, and this was perhaps a little disappointing for him.
The characteristic feature of Carlo's new approach was an essential 
elimination of dynamics. Actually
he was considering just the electromagnetic field, and his main point
was that, even if it be dealt with classically, the electromagnetic 
field has to  contain implicitly  the concept of the photon, an
essential point being that photons necessarily have  to be considered as 
indistinguishable particles. Then, he shows that the 
standard statistical concepts used by Boltzmann, with 
indistinguishability taken into account, leads to Planck's law (the
general Wien's law too, considered as a  property of thermodynamics, should
be taken into account).

Now,  every normal person  would be tempted to say that assuming the
existence of photons is just a direct way of assuming quantum
mechanics itself, and from this point of view Carlo's approach might be considered
to be  just some kind of 
revisitation of the the  Bose--Einstein approach. Everything
depends on the consideration one gives to the way in which
Carlo introduces the photons;  whether this is just some verbal
rephrasing of known things,  or the actual  pointing out of something 
extremely deep, which I'm
still unable to fully appreciate. The extremely high consideration I
have of Carlo suggests to me that this is probably the case, and I'm
looking forward to have an illumination on this point.
In the meantime, just after having written down a first draft of this
paper, I had the opportunity to  pass
an afternoon, sitting in a beautiful garden and  discussing 
this problem with Andrea Carati and Tonino
Scotti, and Andrea has almost convinced us he is on the way of
understanding the actual point made by Carlo. If we will be able to
find some clear statement in this connection, this will be a beautiful
occasion to pay hommage to Carlo, and explain it to everybody.

If we will be able to do this, a key element  will be the following one.
I remember Carlo particularly  pointing  it out to me. The point
concerns the way one should
conceive frequency. We are all accustomed, following Rayleigh, to
think of the electromagnetic field in terms of its normal modes, and thus to
attach  to a packet of normal modes  (or to the
corresponding photons) the label $\omega$ of frequency
 as a parameter, in the same way as
 the mass $m$ is attached as a label to  a particle. However, a particle has also a
   variable velocity $v$, and a corresponding
  kinetic energy $K_m(v)=mv^2/2$. Now, the point Carlo was making to
  me is that, along the lines of Bose,  $\omega$ should be thought of as the analog of $v$,
  rather than of $m$. Indeed,  first
  of all $\omega$ depends on the reference system, as does the velocity of
  a particle,  and even changes within a given reference system when
  light impinges on a moving wall (this actually is a key point in the
  deduction of Wien's displacement law). So the energy $\epsilon$ of a
  particle of light has to be a function of $\omega$, i.e., one should have
 $\epsilon=\epsilon(\omega)$, in the same way in which the kinetic energy of a
particle is a function of velocity.

So if one day, by arguments as those concerning solitons, it will be 
understood that particles of light can be conceived within  
classical physics, then they will have an
energy $\epsilon=\epsilon(\omega)$.  Taking indistinguishability
into account (think again of the interaction with a wall), one then
immediately deduces a Planck--like law involving the
function $\epsilon=\epsilon(\omega)$ of a  still undetermined form. Finally,
Wien's displacement law gives $\epsilon=\hbar \omega$.
Finally,  he adds the zero--point energy as related to ordered, or
nonthermal, motions, and gives an argument (time reversal) to prove that it has
the factor $1/2$ in front.

In any case, whether Carlo's approach will prove to be the good
solution or not, this paper of him is a further witness of the fact
that, even in the last part of his life, he was deeply involved
 in the problem of the relations between quantum and
classical mechanics, with the actual aim of deducing  quantum mechanics from
the classical one.

\section{Some personal recollections about Carlo}
Carlo Cercignani was an extremely reserved person, with whom it was
not easy to have conversations  about deep problems concerning life. 
But he thought very much about that, and his choice was  to commit  his
thinkings to some writings. The first thing he wrote down, that I really
liked very much, is a novel, by the title \emph{``Morte di un
  professore''} or \emph{``Muerte a Pastrufazio''} (he never made a decision
 between such two possible titles). The book is  a joke, as many of the things
he wrote. But chapter 6 was entirely devoted to the verbal
transcription of an ideal talk given by a scientist (by the name
Veroviro, which mimics Truesdell, but the true speaker was Carlo
himself), in which a long discussion is given of the problem whether
liberty of a human being is possible in a deterministic world.

Then he wrote several poems, which were collected in a booklet
\cite{scherzi} by the
title \emph{``Scherzi in versi''}. Among them, the one I like  most is the
poem   by the title 1921 (written in the year 1987), 
which starts with \emph{``Il pendolo semplice e
il verso
  degli angoli, coppie e momenti, e l'orientazion dei segmenti ..''}. I
recited this at his funeral, having made the effort of learning it by
heart.
He also made many very beautiful translations from several poets, as
Shakespeare, Queneau, Borges. He even translated both the Iliad and
the Odyssey, in some verses that he especially chose
with the aim of better fitting the original rhythm of Homer. In the
final years he also wrote one more novel that, according to his wife 
Silvana, is the most beautiful thing he ever wrote. The title is
\emph{``La creazione secondo (according to)  Michele''}.

Just a few months after his passing away, putting some order to my own
bookshelves, I found out a   poem that he gave me perhaps one year
before, and somehow had escaped my attention. This poem seems now particularly
interesting to me, as it gives an indication of how he lived his last
times, when he was almost completely paralized. He actually continued 
to deal with people
in his usual way, somehow as joking at first, then perhaps reciting from
memory with his beautiful voice  a poem suited to the
moment\footnote{One 
of the last times I
  saw him, he  recited  a very beautiful rather long poem by Aleardo
  Aleardi, unknown to me, that   he
  had learned when he was perhaps 13 or 15 years old. In such a poem, a
  description is  given of what goes about when some good people come on a
  battle field after the battle is over, and start  taking care  of the
  bodies of the wounded or dead soldiers.},  then reentering into himself
and starting thinking about something. 

The poem I found has the title 
\emph{``Beethoven in cielo''} (i.e., \emph{in heaven}), and is
essentially a meditation on pain (\emph{dolore}). Carlo
describes how after death he very joyfully goes to heaven. But
strangely enough, the chorus of the angels he  hears is a little
monotonous, and Carlo cannot refrain from  telling them. From the style, he 
recognizes it
as a piece of Beethoven,  but one unknown to him. The angels confirm
it, saying Beethoven composed it in heaven.
 So he asks to meet the composer, what he  is allowed to
do. Requested by Beethoven of his opinion about that music, Carlo at first
refrains to speak, but then, as Beethoven insists,  with
great humility admits  he would have
expected some  more sublime music. And Beethoven says he completely
agrees. \emph{``Tutto in cielo mi vien male, che iattura''}
``Everything in heaven
comes out badly to me. I'm  even  refraining  from composing anymore. You know
why? I'm lacking the creative spark, the note that most shines; this
note is pain ... Only the one who  cries and groans  by pain will have
humanity, divine gift .. Did you ever   cry together with your wife ? 
The one who doesn't do it is unable to capture true love... God too,
 when was seen among us, was he a king, or wanted he to be rich? 
He was  a man's son,
full of pain....'' So Carlo looks at Beethoven, as terrified. ``How strange --
he thinks --  is the flowing of the world. A few hours
ago I was asking  that death should  save  pain to my heart. Now, here,
in this high and blessed (\emph{beato}) world, I regret pain. Oh human 
heart, truly unfathomable and indeed strange.''

This poem is reproduced here in an appendix,\footnote{I thank
  Giancarlo Benettin for typing it for me. In doing this, he could not
refrain from making an almost imperceptible change in  a verse, which was
incomplete.} 
together with the one I
like most, by the title 1921 (1987), that starts with
\emph{``Il pendolo semplice e il verso
  degli angoli, coppie e momenti, e l'orientazion dei segmenti ..''},
and with another, very beautiful, short one by the title \emph{``Leggendo le
  georgiche (1994)''}, that starts with \emph{``Che dir delle stelle, del
  cielo  d'autunno, dell'ansia che prende''}.

\newpage

\textwidth165mm
\textheight248mm 
\oddsidemargin0mm

\voffset=-20mm

\pagestyle{empty}

\obeylines

\def\ii{\`\i}
\def\iis{\`\i\ }
\vskip 3mm
{\bf Beethoven in cielo}
\vskip 3mm

L'anima mia dal corpo si \`e staccata
con una lotta che sar\`a obliata. 
Ma dopo tanta angoscia e sofferenza, 
gioia \`e librarsi come pura essenza! 
II brulichio del mondo \`e come un velo 
e salgo verso un'alta meta: il cielo! 
Lunga e svelta qual freccia \`e la mia via.
Sento lungi una splendida armonia.
D'angeli un coro dolcemente aspetta
per accogliere chi lass\`u si affretta. 
Tra poco sar\`o giunto; quale incanto! 
Che monotono sembri per\`o il canto 
a nascondere agli angeli non riesco.
Ridon felici: -- E' un animo tedesco! 
La musica da voi sale fin qua! 
{\it Dio glorifica\/}, allor, {\it l'eternit\`a\/} 
cantiamo; e veda che ce ne intendiamo!
Ma di andare all'un\ii sono cerchiamo! --
E cantano un corale grande; e presto 
penso: -- E' lo stile di Beethoven questo; 
quel pezzo l\iis per\`o mi \`e sconosciuto. -- 
Chiedo allora: -- Cos'\`e? --  Ordine ha avuto 
di scriverlo -- mi dicon -- dal Signore 
l'anima di Beethoven. Con fervore 
lo eseguiamo ogni volta che c'\`e festa; 
musica qui non c'\`e miglior di questa!--
--Lo credo! Ma vorrei che a me mostrata 
fosse l'anima sua. Non sar\`a stata 
vana cos\iis la gita.-- Divertiti, 
mi conducon per bei prati fioriti 
e m'indican lo spirito divino 
che solitario va, in lento cammino, 
sotto le palme. -- Da lui, bench\'e indegno, 
da lui or voglio andare in questo regno, 
colui che onora pi\`u l'ingegno umano! --
Mi vede allora e mi porge la mano: 
-- Benvenuto, o terreno ospite, prono 
al poter della musica e al suo suono! 
Per te fu il coro angelico eseguito, 
da me composto in cielo, ed ho gradito 
che gli angeli lo affrontin con impegno:
del mio corale, pure in questo regno,
son le quarte eccedenti assai temute!
Ma le mie note, di', ti son piaciute? --
Confuso non rispondo. Ed egli lesto
e cortese prosegue: -- Animo onesto,
tu sei sincero! E' giusta l'opinione:
fuggivi in terra pur l'adulazione. --
Allora del suo dir colgo il vantaggio;
dico: -- O mio eroe -- , facendomi coraggio, 
-- o mio maestro! Ora ho ascoltato il canto
con entusiasmo. Devo dir soltanto
che, tra gli angeli, io, su queste cime,
musica mi aspettavo pi\`u sublime! --
Egli risponde sorridendo: -- Senti,
la penso come te se ti lamenti.
Tutto in ciel mi vien male, che iattura!
Ho smesso di comporre addirittura.
Solo per il giudizio universale
ho dovuto impegnarmi, bene o male,
per non imbarazzare assai il buon Dio,
di scriver per gli ottoni un pezzo mio:
lo devo far, ma non ne sono lieto.
Ma sai perch\'e, altrimenti, ormai mi vieto
di comporre? Mi manca la scintilla
pi\`u creativa, la nota che pi\`u brilla:
questa nota \`e il dolore! S\`\i , il dolore
che ti afferra e ti fa stringere il cuore;
come un metallo forte suona e vibra
e ti fa risuonare in ogni fibra.
E' un vero amico e ti fa superiore:
solo chi piange e geme dal dolore
avr\`a l'umanit\`a, dono divino.
Cosa lega alla madre ogni bambino?
Le grandi pene della notte in cui
c'\`e Dio soltanto in veglia con lei e lui.
Non hai mai pianto insieme con tua moglie?
Chi non lo fa l'amor vero non coglie,
un dolore profondo e condiviso:
il suo ricordo \`e come un paradiso.
Sopporta il santo pena ed afflizione: 
brilla in lui il raggio della perfezione. 
Fama di eroe ottener sol ti \`e concesso, 
se fermamente domini te stesso.
Tremi il tuo cuore nella sofferenza!
Vivrai nel canto della discendenza.
Dio stesso quando qui fra noi si scorse,
fu forse un re, si volle ricco forse?
Fu figlio d'uomo, pieno del dolore,
che ognor s'incontra in qualsisia maggiore
cosa; \`e la nota mia fondamentale.
Ma qui tutto \`e beato e senza male;
la cetra mi \`e caduta allor di mano. --
Lo guardo ora atterrito: -- Come \`e strano
lo scorrere del mondo. Poche ore
fa, chiedevo alla morte che il dolore
al cuore mio venisse risparmiato.
Ora qui, in questo mondo alto e beato,
si rimpiange il dolore! Oh, cuore umano,
veramente insondabile e ben strano!  --
\vspace{2mm}
\hspace{25mm}{\it Carlo Cercignani}

\newpage
 
\vskip 3mm
{\bf 1921 \quad (1987)}
\vskip 3mm

Il pendolo semplice e il verso \ degli angoli, coppie e momenti,
e l'orientazion dei segmenti, \ che sembrano avere un po' perso
quel ruolo di bei caposaldi, \ che noi studiavamo convinti
sui tomi, a caratteri stinti, \ del buon Levi Civita--Amaldi;
i solidi rigidi e i fili \ \ poggiati o sospesi per aria,
disposti per far catenaria, \ e, ancor, coniugati, i profili;
le formule dell'ellissoide \ d'inerzia, le tre rotazioni,
il calcolo delle  reazioni, \ il grafico della cicloide;
e  dopo? anche il moto centrale \ e la geometria delle masse,
e come se ci\`o non bastasse \ le formule del potenziale;
la velocit\`a areolare, \ insieme col noto rapporto
tra assi e periodi e uno storto \ poligono funicolare;
l'epicicloide ordinaria, \ le verghe, i vettori, i versori.
la legge d'inerzia, i cursori, \ l'odografo, la legge oraria;
e quei giroscopi, che ognuno \ ricorda, che sembrano armille;
rinasco, rinasco nel mille, \ eh s\ii, novecentoventuno!
Il Finzi \`e un ragazzo aitante, \ che svolge, sicuro e veloce, 
il calcolo, pure il pi\`u atroce, \ e sgomina ... un determinante;
e ancor la Pastori, serafica, \ in certe serate d'inverno,
risolve, su un lindo quaderno, \ problemi di statica grafica.
Sul tavolo la lucernetta  \ fa un orbe di luce conchiuso,
nel quale discutere \`e d'uso \ problemi di base e rulletta.
Max Abraham,  presso alla morte, \ ancor polemizza ma \`e stanco;
la teppa, rizzata sul banco, \ ne aveva deciso la sorte
da anni: doveva lontano \ andare la \textit{Kultur} germanica
che quel professor di meccanica \ voleva illustrare a Milano.
I giorni si fanni pi\`u foschi; \ che importa se poi il suo tensore
potr\`a risultare migliore \ di quello che ha dato Minkowski?
Ma \`e giunta notizia che oggi \ la vecchia teoria newtoniana,
ed \`e una notizia ben strana, \ su basi sicure non poggi.
Oltr'alpe, se pur tra i fragori \ di guerra, son stati trovati
dei nuovi concetti, e, applicati,\  dei vecchi si trovan migliori.
Qualcuno scuotendo la testa, \ davanti ai propositi empi,
ripete: -- Che tempi! Che tempi!\ \ Dovevo sentire anche questa! --
Tal altro si sente gi\`a certo \ di quella scoperta recente
e dice: -- Che mente! Che mente \ quel tipo, quell'Einstein  ... Alberto! --
Oh sere, passare a studiare \ le formule dell'avvenire,
passate a cercar di capire,\ sapendo pur sol balbettare
\textit{ja, bitte, Forelle e Kartoffel} \quad quei fogli di stampa ancor fresca,
riempiti di lingua tedesca, \ e i simboli di E.B.Christoffel!
Ci sono certuni che stiman \ che sia una pura follia
studiar quella nuova teoria \ con dentro il tensore di Riemann:
che cuore partire all'assalto \ di pagine, quindici a quindici,
ripiene di formule e indici \ che stanno un po' in basso e un po' in alto!
E Weyl, che negli ultimi mesi, \  seguendo i precetti di Mie,
produce ancor nuove teorie, \ chiamate di \textit{gauge} dagli inglesi!
Misteri grandissimi ancor \ ha in serbo la quantizzazione,
descritta con la condizione \ che fu escogitata da Bohr.
Quel mondo si \`e rotto ed ormai \ i giovani studiano a caso
le cose pi\`u strane e col vaso \    Pand\`ora ancor semina guai!
S\iis certi equilibri son rotti \ e circolan libri un po' strani
che srivono, qui, il Cercignani, \ e, un po' pi\`u in l\`a, il Gallavotti.
Perfino ai congressi si sente \ parlare di cose un po' strane:
scompaiono le lagrangiane, \ emerge il fibrato tangente.
O tempo vicino e lontano, \ sei sempre presente nel cuore!
quand'era un versore un versore \ e non un simbolo strano,
ch'\`e simile a una derivata \ parziale, che invece non \`e;
nessuno mi spiega  il perch\'e \ di questa Babele sfrenata.
Eh s\ii, non lo spiega  nessuno \ ed io vorrei che tornasse
quel tempo, che ci si trovasse \ nel milnovecentoventuno.
Tornare nel tempo che fu,\ \  poter imparare i tensori
col Finzi, amar la Pastori! \ quei giorni non tornano pi\`u!
\vspace{2mm}
\hspace{25mm}{\it Carlo Cercignani}

\newpage

\vskip 3mm
{\bf Leggendo le georgiche (1994)}
\vskip 3mm

Che dir delle stelle, del cielo \  d'autunno, dell'ansia che prende
se il sole ogni giorno discende \   pi\`u presto e si copre d'un velo;
o se primavera finisce \ piovosa e nei campi matura
la messe di spighe e Natura \ le steli d'umori arricchisce?
E quando nei campi dorati \ falciare vuoi il fragile orzo
e i venti, rigonfi di sforzo, \ tu vedi scontrarsi adirati
e, come guerrieri nemici, \ avvolti  di nuvole scure,
strappare le spighe mature, \ svellendo perfin le radici,
scagliarle nell'aria; e le nere \ tempeste avvolgere in spire,
facendoli in alto salire, \ gli steli e le stoppie leggere.
Il cielo non \`e pi\`u celeste \  e senti continua scrosciare
la pioggia; salita dal mare, \ la nube si addensa in tempeste
oscure d'orribile pioggia; \  il cielo precipita in terra
e scende feroce a far guerra. \   Si gonfiano il fiume e la roggia,
con strepito il mare ribolle; \ si allagano i campi ridenti
dell'opra di giorni pazienti \ e sul seminato e sul colle.
Furente con noi Giove Pluvio, \ brandendo la folgore iroso
ci manda col buio nuvoloso, \ o sembra, un suo nuovo Diluvio...

\vspace{2mm}
\hspace{25mm}{\it Carlo Cercignani}

\end{document}